\documentclass[11pt,twoside]{article}

\usepackage{asp2006_title}

\usepackage{graphicx}

\newcommand{\gr}{$\gamma$-ray}
\newcommand{\grs}{$\gamma$-rays}
\newcommand{\pz}{$\pi^{\rm o}$~}
\newcommand{\pzd}{$\pi^{\rm o}$-decay}

\begin{document}
\title{Gamma-rays from star-forming regions:\\
 from SNOBs to dark accelerators}   
\author{Thierry Montmerle}   
\affil{Laboratoire d'Astrophysique de Grenoble, France}    

\begin{abstract} 
Observational \gr~astronomy was born some forty years ago, when small detectors were flown in satellites, following a decade of theoretical predictions of its potential to discover the origin of cosmic rays via the \pzd~mechanism. The seventies were a golden era for \gr~and cosmic-ray astrophysics, with the (re)discovery of the ``diffuse shock acceleration'' theory for cosmic rays, and the first CO and GeV \gr~surveys of the galactic plane, verifying the importance of \pzd~in the large-scale \gr~emission of the Galaxy. But because of this strong galactic background, GeV \gr~sources were hard to identify. The first such sources definitely identified were three pulsars, with a suggestion that supernova remnants interacting with molecular clouds in massive star-forming regions (``SNOBs'') were also \gr~sources. Because of their improved sensitivity and spatial resolution, ground-based \u{C}erenkov telescopes, detecting \grs~at $>$ TeV energies, are now able to resolve molecular cloud-sized objects at a few kpc. SNOB-like objects like IC443 and W28 are detected at GeV and TeV energies, and show spatial evidence for cosmic ray interactions between an SNR shock wave and nearby molecular clouds, and subsequent \pzd. However, the spectral evidence does not clearly support this mechanism. We propose to use another tool for probing the interaction of the low-energy component of the putative local cosmic rays, in the form of enhanced ionization in TeV-bright molecular clouds, using millimeter observations.
\end{abstract}


\section{Historical context: the ``roaring seventies'' of high-energy astrophysics}  

Gamma-ray astronomy and cosmic rays have a long history in common. In the early sixties, it was recognized that the decay of \pz mesons into 2 $\gamma$-rays (discovered at the Berkeley cyclotron in 1950) could be a major mechanism for producing $\sim 100$ MeV $\gamma$-rays in the Galaxy as a result of collisions between high-energy ($> 1$ GeV) cosmic-ray protons and hydrogen atoms in the interstellar medium. The first predictions were published in physics journals, soon after the experimental discovery (Hayakawa 1952), with the ultimate goal to unravel the origin of cosmic rays, in particular their potential link with supernova remnants, both as energy providers and acceleration sites (e.g., Pollack \& Fazio 1963, Hayakawa et al. 1964, Ginzburg \& Syrovatskii 1964; Stecker 1970, etc.). It took however several years before $\sim 100$ MeV $\gamma$-rays were actually detected as diffuse emission coming from the Galaxy (discovered with a 10-cm diameter scintillation counter aboard the OSO-3 solar satellite: Clark et al. 1968, Kraushaar et al. 1972), with possible localized enhancements (Cygnus region, ballon experiments, Niel et al. 1972). At the same time already, TeV $\gamma$-ray emission was detected from the Crab pulsar by \u{C}erenkov telescopes, thanks to its pulsations, coming not from  \pz decay but from synchrotron emission  (Grindlay 1972).

The real observational breakthrough came with satellites fully dedicated to $\gamma$-ray detection in the range $\sim 10$ MeV to $\sim 1$ GeV, via the detection of e$^+$e$^-$ pairs in $\sim 20$-cm diameter spark chambers: {\sl SAS-2} (NASA, 1972-1973), and {\sl COS-B} (ESA, 1975-1982). Whereas the reconstruction of the electron tracks provided some directional information about the incoming $\gamma$-rays, the impossibility to measure the recoil of the heavy nucleus {\it Z} involved in the reaction $\gamma$ + {\it Z} $\rightarrow$ e$^+$ +  $e^-$ + {\it Z} limited the angular resolution to $\sim 1\deg$ at best. This was enough to claim the detection of $\gamma$-ray ``sources'' (as ``hot spots'' of emission over some local background; see below), but not quite enough, in the galactic plane and in the absence of timing information, to claim a secure identification with an astronomical object. Owing to the premature failure of its tape recorder to store and transmit the data on the ground, {\sl SAS-2} had the time to provide only a glimpse of the galactic plane in $\gamma$-rays, but discovered the pulsed emission of the Crab and Vela pulsars (Fichtel et al. 1975). In contrast, {\sl COS-B} succeeded, in 7 years of operation, to deliver the first all-sky map in $\gamma$-rays in the 70 MeV-5 GeV range (``GeV $\gamma$-rays'' for short in what follows), with particular emphasis on the galactic plane (Mayer-Hasselwander et al. 1982).

The interpretation of the galactic GeV emission was made possible with two other, almost simultaneous, major developments: $(i)$ the recognition of the CO molecule as a proxy for (cold) molecular hydrogen, impossible to observe in its ground state for lack of polarity, and the subsequent first CO survey of the galactic plane in mm waves (with a 1.2-m diameter telescope built on a roof at the University of Columbia, in uptown New York city !); $(ii)$ sudden progress in the theory of cosmic-ray acceleration in the form of the so-called ``diffusive shock acceleration'' theory, based on the early ``Fermi type-I'' mechanism (Fermi 1949).

$\bullet$ {\it Comparison between CO and GeV $\gamma$-ray maps of the galactic plane.} As the {\sl COS-B} and CO surveys were progressing, it became increasingly clear that, to first order, the original predictions of $\gamma$-ray emission on the galaxy were correct -however including now molecular hydrogen, unknown at the time, as the main component of mass in the galaxy. As illustrated in Fig.1, the similarity between the maps (initially at comparable resolutions: 0.5$\deg$ for CO, see Dame \& Thaddeus 2004 for a modern reference, $vs$. $\sim 1\deg$ for GeV $\gamma$-rays) was striking, implying a tight correlation between molecular clouds and $\gamma$-ray emission. This had three consequences: $(i)$ the dominant $\gamma$-ray emission is \pzd \footnote{Other processes, linked with the high-energy {\it electron} component, i.e., bremsstrahlung and inverse Compton, turned out to be negligible on a galactic scale} ; $(ii)$ the distribution of cosmic-rays in the galaxy is almost uniform, implying a fast diffusion of particles away from their sources -only a slight arm-to-interam ratio of a factor of $\sim 2$ in cosmic-ray density could be perceptible; $(iii)$ put in another way, and perhaps unfortunately, there are no obvious ``sources'' of cosmic rays that would produce bright hot spots in GeV $\gamma$-rays. On the contrary, at GeV energies, the galaxy appears as a bright, extended but complex region of diffuse emission (see Casanova et al. 2009),  which is the main background against which any potential ``source'' (like a supernova remnant) has to be extracted.


\begin{figure}[h]
\begin{center}
\includegraphics[width=13cm]{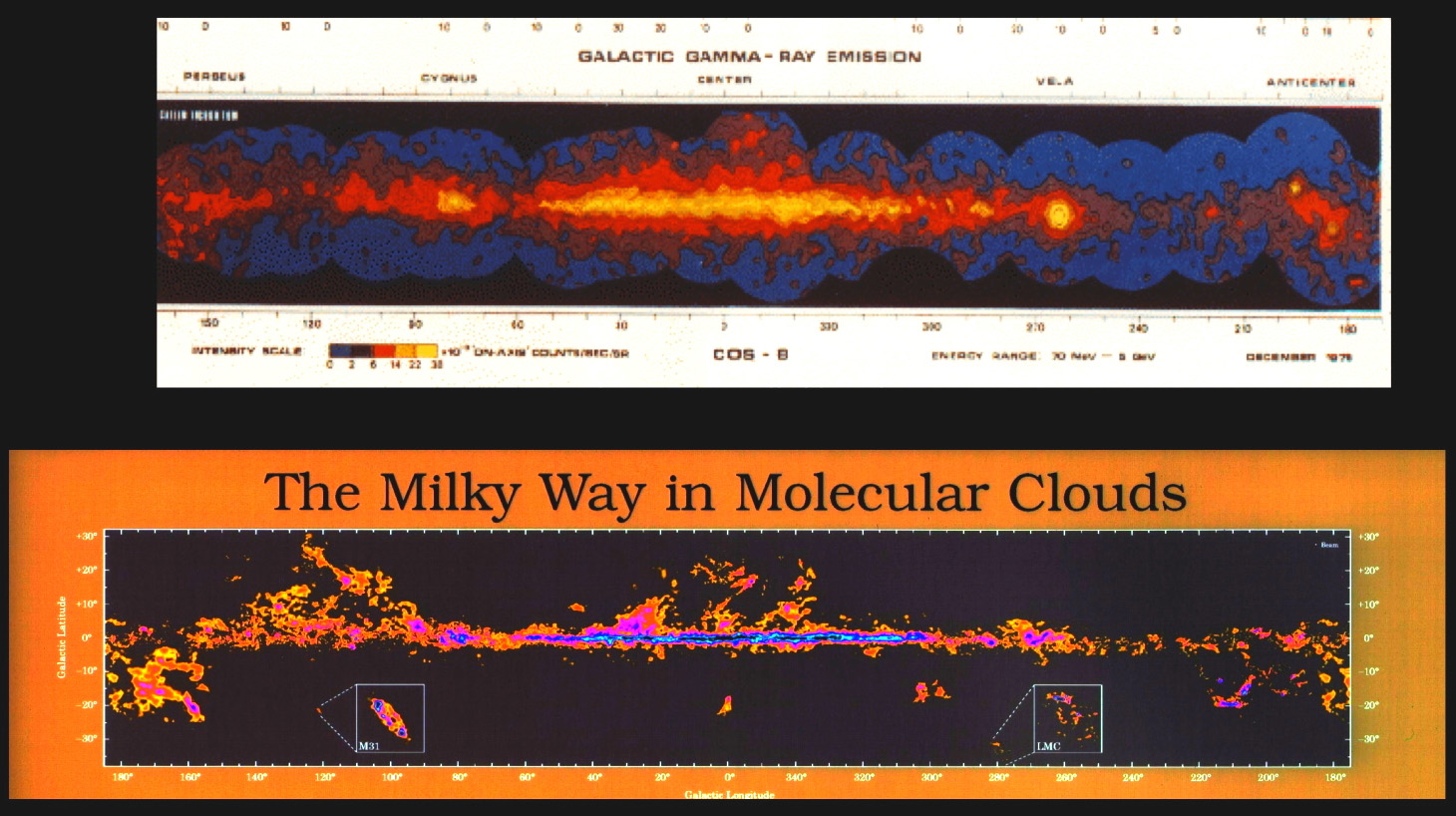}
\end{center}
\caption{The galactic \gr~emission at $\sim$ GeV energies revealed by {\sl COS-B} (from Mayer-Hasselwander et al. 1982, top), aligned with the corresponding $^{12}$CO emission (from Dame et al. 1981, bottom)}
\label{fig.1}
\end{figure}


$\bullet$ {\it Cosmic-ray acceleration theory.} In the late forties, E. Fermi proposed that cosmic rays could be accelerated by a combination of two physical processes: $(i)$ coupling of (ionized) cosmic-ray nuclei with interstellar magnetic fields; $(ii)$ relative motions of interstellar clouds, themselves tied to magnetic fields and thus generating magnetic fluctuations (Alfv\' en waves), allowing exchanges of energy with cosmic rays by a ``bouncing'' process between the interstellar clouds. Fermi showed that a net positive energy input into cosmic rays would result from the stochastic motions of the interstellar clouds. His original equation for the energy gain (Fermi 1949) can be expressed as:
\begin{equation}
\frac{\delta E}{E} = 2 \frac{v v_B}{c^2}/(1 - \frac{v^2}{c^2})
\end{equation}
where $E$ is the energy of cosmic rays of velocity $v$, and $v_B$ is the velocity of the magnetic perburbations induced by the mutual cloud movements: the energy grows basically exponentially.

In the late '70's, several authors independently realized that the same mechanism would -perhaps more realistically- operate between the upstream and downstream regions of a shock wave propagating into the interstellar medium (Axford et al. 1977, Bell 1978a,b, Blandford \& Ostriker 1978, Eichler 1978). In brief, for an adiabatic shock, the ratio of upstream ($u_1$) to downstream ($u_2$) velocities $p = u_1/u_2$ depends on the energy exchanges between the shock and the surrounding medium (Rankine-Hugoniot's equations; $p = 4$ for an adiabatic shock); the gas is compressed upstream, and so is the magnetic field, implying a sudden change in the Larmor radius of cosmic rays crossing the shock, and a gain in energy. Particles bounce off magnetic field fluctuations (themselves traveling at the Alfv\' en velocity $v_A$) by resonant scattering, i.e., diffuse back and forth on both sides of the shock, with a net increase of energy each time they enter the upstream flow. Compared to the initial Fermi mechanism, the ``diffusive shock acceleration'' mechanism (or DSA) as it was soon called, had one enormous additional advantage: from the start, it explained in a relatively ``natural'' way the slope of the cosmic-ray spectrum above $\sim 1$ GeV, i.e. $\alpha \sim 2.5$ with $N(E) \propto E^{-\alpha}$. For instance, in its simplest form (e.g., Bell 1978a), on finds $\alpha = (p+2)/(p-1), = 2$ for $p=4$. Taking into account the shock velocity $v_s$ and  the movement of the scattering centers (with velocity $v_w << v_s$), one finds a slightly more complicated equation:
\begin{equation}
\alpha=\frac{(p+2)+p(2v_w/v_s-v_A/v_s)}{(p-1)-p(v_w/v_s+v_A/v_s)}
\end{equation}
The observed value $\alpha = 2.5$ can be obtained for an adiabatic shock if $v_w=0.08$, or alternatively, if $v_w=0$, with an Alfv\'enic Mach number $v_s/v_A \sim 4$, giving $p \sim 3$ rather than 4, consistent with some measurements in interplanetary space.
However, it was soon recognized that the proposed new mechanism was in fact too efficient: it could tap the shock energy so much that it would, in fact, quench the shock itself, so that $p \rightarrow 0$. In other words, the back-reaction of the accelerated particles on the shock had to be taken into account -and the price to pay was that the spectral slope of the cosmic rays could not be easily explained any more... And other effects, such as the orientation of the shock with respect to the magnetic field, radiative losses, etc., also had to be taken into account.

Thirty years after, in particular with the increased power of numerical simulations, the DSA is still the leading  mechanism for cosmic-ray acceleration, but much more sophisticated and altogether in a much better shape; in particular, many details of the theory could be tested against supernova observations (e.g., Ellison et al. 2007) or  in situ observations of interplanetary shocks (see also Ellison, this meeting).

\section{GeV $\gamma$-ray sources and the SNOB hypothesis}

The first {\sl COS-B} catalog of $\gamma$-ray sources (Hermsen et al. 1977), covering $\sim$ 1/3 of the galactic disk, contained 13 entries. The only unambiguous sources were the Crab and Vela pulsars, with their conspicuous double-pulse light curves. Another source, seemingly genuinely pointlike because it lied away from any CO detection, but lacking any obvious counterpart, was dubbed {\it Geminga}.{\footnote{A pun between ``Gemini gamma-ray source'' and the nickname given by G. Binami and his Milanese colleagues of the {\sl COS-B} collaboration: ``Gemin g'ha'', which means ``There's nothing'' in the local dialect.}} This source, which necessitated over twenty years of stubborn multiwavelength observational work to be eventually identified with the closest pulsar known ($d \sim 150$ pc) (e.g., Salvati \& Sacco 2008), may be considered as the first example of a ``dark accelerator'', long before that name would be given to TeV sources without counterpart (or with too many !; see \S 6)...

Ten GeV sources remained unidentified. They were defined as localized excesses of $\gamma$-ray flux, found after modelling, as accurately as possible using the CO data available at the time, the expected galactic $\gamma$-ray diffuse background emission. Then Montmerle (1979) noted that, for over half of these unidentified sources, the $\sim 1\deg$ diameter error box included a shell-like structure, generally a known supernova remnant, associated with a cluster of massive stars (``OB associations'') and giant molecular clouds, hence the nickname ``SNOBs'' given to such sources. The key idea was based on the concept of ``propagating star formation'' introduced shortly before by Elmegreen \& Lada (1977): massive stars evolve in supernovae in a timescale short enough (a few Myr) that they explode essentially on the site of their formation (that would be the SNOB phase), and their shock wave may then interact with the parent molecular cloud, trigger the birth of a new stellar association, and at the same time accelerate cosmic rays.

As far as GeV emission was concerned, the SNOB concept, and its later generalization to high-speed winds from massive stars (e.g., Cass\' e \& Paul 1980, Cesarsky \& Montmerle 1983),{\footnote{Winds from massive stars were also discovered in the '70's; see Kudritzki \& Puls (2000) for a review}} contained all the basic ingredients to make a convincing $\gamma$-ray source: powerful shock waves to accelerate particles by the DSA mechanism, a large amount of mass in the form of giant molecular clouds as targets for high-energy particle collisions, and subsequent \pz creation and decay. It even had a wealth of low-energy particles around to inject into shock waves, from strong flares from the low-mass stars of the same association, discovered in X-rays shortly thereafter (e.g., Feigelson \& DeCampli 1981, Montmerle et al. 1983).

The proposed identification of a number of {\sl COS-B} sources with SNOBs or SNOB-like objects had a number of immediate implications:

- Knowing the target mass available in the form of HII regions and/or molecular clouds, typically in the range $\sim 10^5 - 10^6 M_\odot$, using the \pzd~GeV $\gamma$-ray emissivity (e.g., Stecker 1970, 1971), one finds that the local energy density of high-energy cosmic-rays $\eta_{CR}$ irradiating the clouds must be up to $10-100$ times higher the average GCR energy density ($\sim 1$ eV~cm$^{-3}$);

- However, the inelastic collision cross section is $\sigma_{pp,inel} = 3 \times 10^{-26}~{\rm cm}^2$ (e.g., Mannheim \& Schlickeiser 1994), corresponding to a mean free path of high-energy protons, expressed in g~cm$^{-2}$ as is customary in cosmic-ray physics, of $\sim 50~{\rm g~cm}^{-2}$ at 1 GeV. Expressed in usual astronomical column density units, this is equivalent to $N_H \sim 3 \times 10^{25} ~{\rm cm}^{-2}$, or $A_V \sim 2000$ magnitudes of extinction (using, e.g., Vuong et al. 2003). This number can be compared with the typical column density of molecular cores, $N_H \sim$ a few $10^{22} {\rm cm}^{-2}$, or $A_V \sim 50$. This implies that the high-energy cosmic rays must be efficiently {\it confined} within the molecular clouds. This can conceivably be done by turbulent magnetic fields within the HII region or within the clouds. Zeeman observations indicate that in HII regions magnetic fields are on the order of $B \sim 10-20~\mu$G (Heiles et al. 1981); in molecular clouds, $B \sim 1-100~\mu$G, with a strong correlation between $B$ and the gas density $n_H$ (e.g., Troland \& Crutcher 2008). If the cosmic rays are not efficiently confined, their energy density must be correspondingly higher.

Well-studied SNOBs include the IC443 and W28 supernova remnants, which we shall discuss below (\S 4).

\section{From GeV to TeV \gr~sources}

After {\sl COS-B} ended operations, it took almost ten years to see another \gr~satellite in orbit: the {\sl Compton Gamma-Ray Observatory}, aka CGRO (NASA, 1991-2000). Among other instruments, it carried the ``EGRET'' two-stage square spark chamber, an improved, much larger ($55 \times 55$~cm)  version of its predecessors, operating in a more extended energy range (20 MeV to  30 GeV) and with an improved angular resolution ($\sim 5-30'$, depending on the arrival direction and energy of the incoming $\gamma$-rays). Like {\sl COS-B}, CGRO/EGRET could establish in its nine years of operational lifetime an all-sky map in $\sim$ GeV \grs, in particular of the galactic plane, as well as a catalog of nearly 300  $\gamma$-ray sources: the ``Third EGRET Catalog'' (3EG) contained 271 sources (Hartman et al. 1999). However, in large part due to inaccuracies in modelling the galactic diffuse background (even at intermediate galactic latitudes), this catalog was shown later to necessitate many revisions: Casandjian \& Grenier (2008) reach a total of 188 sources, with the suppression of 107 3EG sources, but with the addition of 30 new ones.

Even with these modifications, the EGRET results demonstrated that SNOB-like objects were confirmed to be a class of GeV \gr source (see Torres et al. 2003 for a detailed review).

Successor to CGRO, the {\sl Fermi} (ex-GLAST) satellite was launched in June 2008. Its carries the LAT ({\sl Large Area Telescope}), which operates in the 30 MeV-300 GeV range, This instrument is 30 times more sensitive than EGRET and has an angular resolution of $\sim 1'$. It is based on an entirely new spark-chamber array design, derived from particle physics. The LAT is expected to give an entirely new view of the GeV sky. However, notwithstanding its spectacular performance, it will have to face the same problems of identifications of GeV \gr~sources in the galactic plane, i.e., a strong galactic background and its modelization (including cosmic-ray electrons; see Porter et al. 2007). The problem of the comparison with the distribution of molecular hydrogen is reversed, in that the LAT angular resolution is now {\it better} than that of large-scale CO surveys (Columbia: $1/8\deg$: NANTEN (Nagoya): $\sim 3'$). So only ``strong'' galactic GeV sources, with a high contrast (or with time-resolved light curves, of course) will be available at first.

Pending the {\sl Fermi} results, the most significant breakthrough has been recently accomplished not from space, but on the ground, with the coming of age of \u{C}erenkov telescopes, after decades of frustrating research, with little more than the detection of the Crab pulsar (discovered in the '70s as mentioned above) as a result. \u{C}erenkov telescopes detect the atmospheric UV radiation emitted by extensive air showers of secondary particles generated when a very high-energy particle or nucleus hits the upper layers of the atmosphere, at an altitude of $\sim 20$ km. The incoming particles have enormous energies, in the $\sim$ 100 GeV-100 TeV range (hereafter the ``TeV range'' for short).

Thanks to the new generation of \u{C}erenkov telescopes, and in particular the HESS 4-telescope array in Namibia, operational since 2004, ``TeV astronomy'' has become a reality. The detection of several dozen sources so far has been reported, covering a wide variety of celestial objects: galactic like pulsar wind nebulae, high-mass X-ray binaries, and supernova remnants; extragalactic like AGNs or blazars. 

Even if one takes into account the differences in sky coverage between  \u{C}erenkov and EGRET observations, the GeV and TeV skies are spectacularly different. In particular, forgetting about the extragalactic sources, Fig. 2 shows a greater concentration of TeV sources along the galactic plane compared to GeV sources.


\begin{figure}[h]
\begin{center}
\includegraphics[width=10cm]{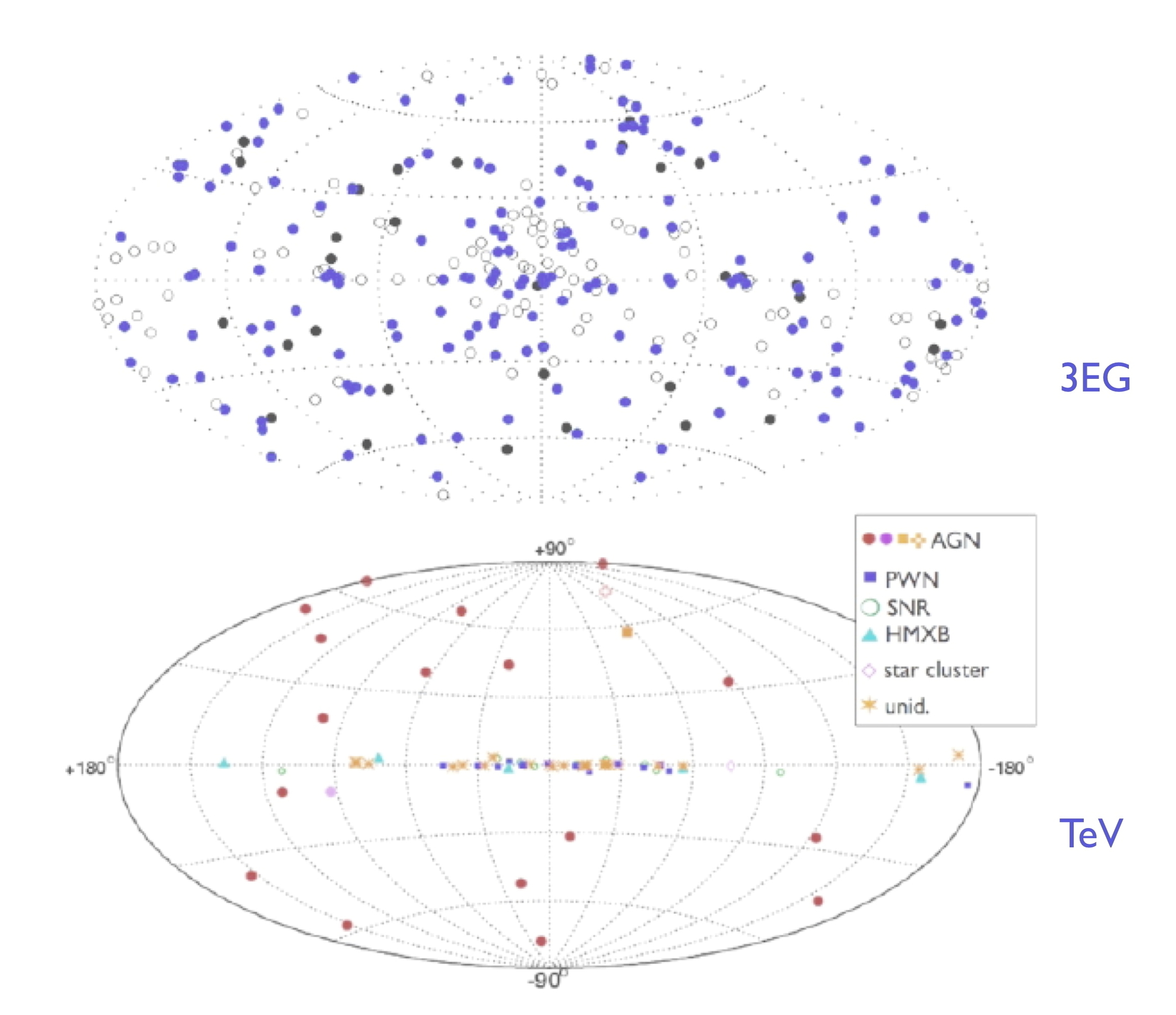}
\end{center}
\caption{Comparison between the GeV sky (3rd CGRO/EGRET catalog, top), and the current TeV sky observed by \u{C}enrenkov telescopes (bottom) (Grenier 2008)}
\label{fig.2}
\end{figure}


This is the result of a contrast effect: the Milky Way is very bright at GeV energies, but existing \u{C}erenkov observations (for instance the HESS survey of the galactic plane, Aharonian et al. 2008a) do not show evidence for a significant background emission at TeV energies. This may be due in part to the slope of the galactic cosmic-ray spectrum and the slope of the resulting \pzd~\gr~emission, but for the moment it is likely the consequence of the background suppression method over the large field of view of \u{C}erenkov telescopes. In any case, the consequence is that, somewhat paradoxically, galactic TeV sources turn out to be much more ``visible'' than galactic GeV sources. They also appear to represent different populations, and so far only a few sources are seen both at GeV and TeV energies. However, this may be an observational bias, and future {\sl Fermi} results may well change this conclusion.

\section{SNOBs as TeV sources ?}

Among current joint GeV(EGRET)-TeV sources, one can find a few SNOB-like sources, such as IC443 and W28. We will briefly discuss them in turn, to illustrate how they may shed a new light on the problem of the origin of cosmic rays. However, it is not yet clear that, in these sources, the \gr~emission (both at GeV and TeV energies) results from \pzd, following collisions between high-energy ($>$ GeV) cosmic rays (mainly protons), locally accelerated by the SNR shock, and molecular clouds in their immediate vicinity, or from an electron process intrinsic to the SNR (like synchrotron emission) since the \gr~spectra are also consistent with power-laws.


\begin{figure}[h]
\begin{center}
\includegraphics[width=12cm]{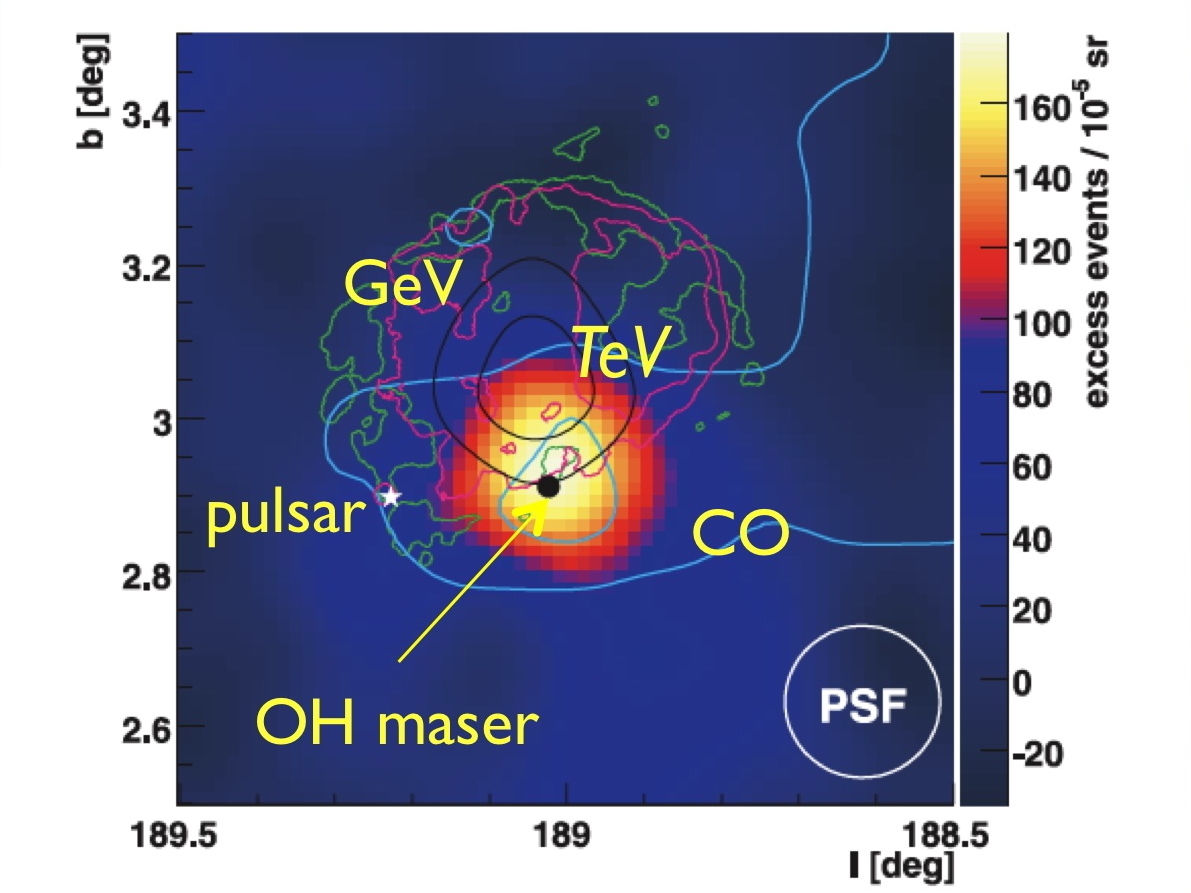}
\end{center}
\caption{The IC443 SNR and its environment (based on Fig. 1 of Albert et al.  2007). The familiar H$\alpha$ shell is outlined (wiggling contours), along with the CO cloud (smooth outline), the OH maser, and the pulsar wind nebula. The EGRET GeV source contours, and the MAGIC TeV source, are also indicated.}
\label{fig.3}
\end{figure}


$\bullet$ {\it IC443.} This famous supernova remnant (nicknamed ``the medusa''), in the Gemini constellation, is close to the IC444 nebula. At a distance $d\sim 2$ kpc, it is a comparatively old SNR, with an estimated age of $\sim 3 \times 10^4$ yrs. In the radio, the non-thermal emission appears essentially circular. In X-rays, it belongs to the ``center-filled'' morphology, in other words the hot plasma does not fill the radio contours. A spectacular (but small in size compared with the SNR) ``pulsar wind nebula'' has been discovered in X-rays, near the outer shell of the SNR and definitely off-centered (see references in Albert et al. 2007a). There is ample evidence that the SNR shell is interacting with dense material: the existence of an OH maser indicate that the shell is actually colliding with dense material, and several indicators  give a shock velocity $v_{sh} \sim 100~{\rm km~s}^{-1}$ (Hirschauer et al. 2009). Also, a detailed {\sl XMM-Newton} study of the whole SNR shows that it is in reality made of two opposite shells of slightly different sizes, squeezed by a sausage-shaped molecular cloud (Troja et al. 2006), which is detected in CO. From the respective locations of the OH maser and of the CO cloud, and the X-ray extinction, it seems  that the interacting part of the shock wave is propagating mainly towards the observer.


\begin{figure}[h]
\begin{center}
\includegraphics[width=10cm]{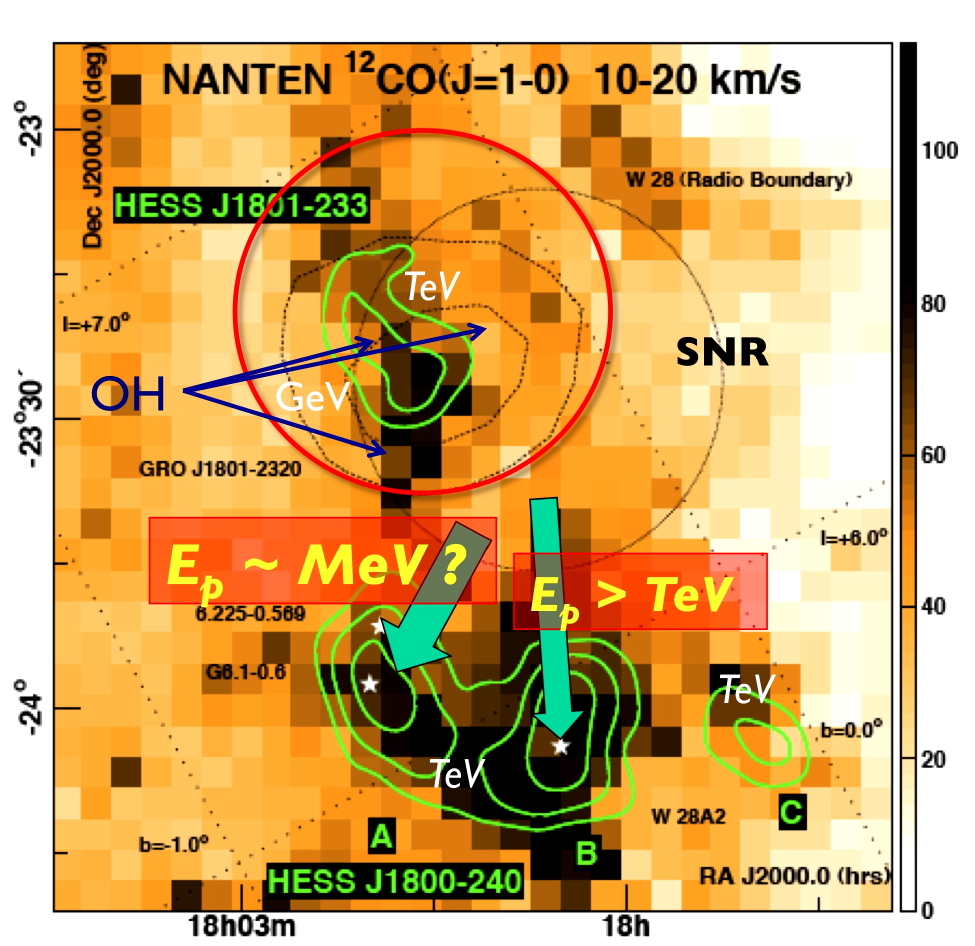}
\end{center}
\caption{The W28 SNR and its environment (based on Fig. 2 of Aharonian et al. 2008b). The background squares correspond to $^{12}$CO intensities. Note the position of the TeV-bright CO clouds away from the SNR shock, strongly suggesting that they are irradiated by $\sim$ TeV cosmic rays accelerated by the SNR. If $\sim$ MeV particles can also diffuse out to the clouds, enhanced ionization will result, which can be measured with millimeter telescopes }
\label{fig.4}
\end{figure}


$\bullet$ {\it W28.} This SNR lies within a complex star-forming region in Sagittarius, almost exactly along the galactic plane, not very far from the M20 (Trifid) nebula. Estimates for its distance and age vary between 1.8 and 3.3 kpc, and 3.5 and $15 \times 10^4$ yrs, respectively. It is also known in the radio cm domain, where it dominates a wide bell-shaped emission region comprising several components. In X-rays, it is also a ``center-filled'' source (see references in Aharonian et al. 2008b). Several OH masers are detected near the radio cm boundary (Claussen et al. 1999), and shock velocities $v_{sh} \sim 70~{\rm km~s}^{-1}$ are found (Mavromatakis et al. 2004).   CO observations reveal the presence of several molecular clouds in the vicinity of the SNR. The respective positions of the OH masers and of the CO clouds indicate that the shock interaction takes place more or less in the plane of the sky.

Both IC443 and W28 have been detected at GeV and TeV energies, and it is interesting to compare them. Fig. 3 shows the environment of IC443, locating the EGRET GeV and MAGIC TeV sources, as well as the main features depicted above. Perhaps the most remarkable fact is that the pulsar associated with the SNR is not detected at high energies -whereas pulsar wind nebulae make up the majority of HESS sources along the galactic plane. On the other hand, the TeV source appears unresolved, and its position is coincident with that of the OH maser, and, within errors, with the GeV source. In particular, the TeV emission does not appear to come from the bulk of the molecular cloud as traced by CO.

Fig. 4 displays the same basic features as IC443, but with some important differences. $(i)$ Judging from the spatial distribution of the OH masers, the SNR shock interaction takes place more or less in the plane of the sky; $(ii)$ the HESS TeV emission is extended, comprising four resolved components well correlated with CO clouds; $(iii)$ only one TeV component is associated with an EGRET GeV source; this component lies in the area where the OH masers are found, i.e., of maximum shock interaction; $(iv)$ the three other TeV components lie at some distance from the SNR southern boundary (about $0.5 \deg$, or $ \sim 3.5$ pc taking a distance $d \sim 2.5$ kpc at least in projection), but none of them seems associated with GeV emission.

Thus, (over)simplifying, comparing IC443 and W28 as observed from the Sun, the two SNRs can be considered roughly as two faces of the same object, namely a SNR interacting with a molecular cloud, barring some adverse projection effects: the IC443 shock is interacting with a molecular cloud located mostly around its ``equator'', and W28 shows a similar interaction, but seen ``pole on''. Then IC443 would resemble W28 seen, in the plane of the sky, from $45 \deg$ to the NE in Fig.4, i.e., where the GeV and TeV \grs, and OH masers, are simultaneously detected. In W28 at least, the tight spatial correlation between TeV and CO emissions is strongly indicative that the \grs~come from \pzd, in other words, from locally accelerated protons colliding with molecular material. However, in both cases the GeV-TeV \gr~spectrum is also consistent with a power-law, so that alternative interpretations are not ruled out (see discussions in Albert et al. 2007a and Aharonian et al. 2008b, respectively). 

So even in these two well-studied cases, the \gr~emission mechanism is not well established. A power-law spectrum at TeV energies has been found in isolated, young SNRs like Cas A (Albert et al. 2007b), i.e. produced by very high-energy {\it electrons}, either/or by synchrotron or inverse Compton mechanisms. If the \gr~spectra of IC443 and/or W28 are power-laws, this would indicate that only the {\it leptonic} emission of the SNR is detected. If, on the other hand, the \gr~spectra are from \pzd, then it will be an observational proof, at last, that {\it hadronic} interactions take place and that, after all, we have evidence for proton acceleration by SNR shocks.

Future \gr~observations, with better statistics, may resolve the problem. But since, as argued above, \pzd~\gr~emission implies a deep penetration and confinement of cosmic-ray protons within the molecular cloud, another method may be used: probing for enhanced ionization in the ``\gr~bright'' molecular clouds, by the {\it low-energy component} of the cosmic rays accelerated by the SNR shock. We discuss this point in the next section.

\section{Cosmic-ray irradiated molecular clouds: the ionization test}

Therefore, let us assume, from now on, that locally accelerated cosmic rays (``LCR'' hereafter), with an energy density one to two orders of magnitudes higher than galactic cosmic rays (GCR), are responsible for the GeV and/or TeV emission of molecular clouds in SNOB-like objects. As is well known, the currently observed range of ionization rates in dense molecular clouds ($\zeta \approx 10^{-17\pm 0.5}~{\rm s}^{-1}$, Caselli et al. 1998) is generally attributed to GCR protons and nuclei of energies $< 100$ MeV/n, although recent calculations show that the electron contribution, in the same energy range, is probably important (Mannheim \& Schlickeiser 1994; Padovani et al. 2009). To first order, depending on the LCR spectrum (which depends on the local cosmic-ray propagation, see below), the ionization rate can be expected to increase in the same proportions as high-energy LCR ($E_p > 10$ TeV, say), and measured, at least relatively to neighbouring, non-\gr~emitting molecular clouds.

Different methods can be used to measure the ionization degree, $x_e = n_e/n_H$ of molecular clouds, which is related to the ionization rate by $x_e = \zeta/(\beta n_H)$, where $\beta \sim 10^{-7}~{\rm cm}^3{\rm s}^{-1}$ is the recombination rate in dense molecular clouds (e.g., Glassgold et al. 2000). For instance, one can use the DCO$^+$ and HCO$^+$ radicals, created via the cosmic-ray ($CR$ = low-energy protons or electrons) sensitive reactions (e.g., Oka 2006):\\

$CR$ + H$_2$ (HD) $\rightarrow$ H$_2^+$ (HD$^+$) + $e^-$ + $CR'$ 

H$_2^+$ (HD$^+$) + H$_2$ $\rightarrow$ H$_3^+$ (H$_2$D$^+$) + H\\

CO + H$_3^+$ $ \rightarrow$ HCO$^+$ + H$_2$

H$_3^+$ + HD $ \rightarrow$ H$_2$D$^+$ + H$_2$

H$_2$D$^+$ + CO $ \rightarrow$ DCO$^+$ + H$_2$\\

Then DCO$^+$/HCO$^+$ = $\frac{1}{3}$(H$_2$D$^+$/H$_3^+$)/[$1 + \frac{2}{3}$(H$_2$D$^+$/H$_3^+$)]. Also, it can be shown that in standard
molecular clouds the H$_2$D$^+$/H$_3^+$ ratio depends on the ionization degree
 as follows (Gu\'elin et al. 1977; Ceccarelli \& Dominik 2005):
\begin{equation}
  \label{eq:h2d+}
  \mathrm{%
\frac{H_2D^+}{H_3^+}}=
\frac{2 \cdot [\mathrm{D}] k_{1}}{%
k_\mathrm{e}x_\mathrm{e} + 
k_\mathrm{CO}x_\mathrm{CO} +
2k_{HD}[\mathrm{D}]
}
\end{equation}
\noindent where $k_{HD}$, $k_e$ and $k_{CO}$ are the rate coefficients of the
reactions of H$_2$D$^+$ with HD, electrons and CO respectively, $x_e$
and $x_{CO}$ are the electronic and CO abundances, and [D] is the
elemental abundance of deuterium relative to H nuclei,
$D/H = 1.5\times10^{-5}$ (Linsky 2003); $k_{1}$ is the rate coefficient of
the reaction of H$_3^+$ with HD which forms H$_2$D$^+$. 

At the same time, the cloud mass, to which the \gr~emission should be proportional if high-energy LCR fully interact with it, can be obtained from $^{12}$CO and $^{13}$CO measurements.

We (Ceccarelli, Montmerle, et al.) are currently engaged in such observations of IC443, W28 and other HESS sources, using various millimeter telescopes. The ultimate goal is to check whether low-energy (10-100 MeV, ``MeV'' for short in what follows) LCR do interact also with the cloud, in other words whether the MeV protons and nuclei accelerated by the SNR shock have had the time to diffuse out to to the cloud like TeV ones do. To illustrate the point, let's take the case of W28. Apparently, in spite of comparable masses, the TeV-bright molecular clouds are not detected at GeV energies, which may mean that $\sim$ GeV LCR have not had the time to reach the clouds. Then, taking the usual energy dependence of the GCR diffusion coefficient ($D \propto E^{0.5}$) (see Gabici et al. 2009), MeV LCR would be even less likely to interact with the cloud and ionize it to levels higher than the usual GCR values. Once inside the cloud, the LCR ionization enhancement would be $\sim 30 \times \lambda_{CR}/10{\rm pc}$, where $\lambda_{CR}$ is the diffusion mean free path (Fatuzzo et al. 2006). Thus, depending on the $x_e$ value found from our observations, we expect to find a strong constraint on the propagation of LCR from the shock to the dense parts of the clouds.

\section{``Dark accelerators''... ?}

The current census of TeV \gr~sources along the galactic plane suggests that most of them are linked with the last stages of stellar evolution: pulsar wind nebulae and supernova remnants (interacting or not with molecular clouds), or X-ray binaries. But to date many sources remain unidentified, i.e., do not seem to have a well-defined counterpart, including in the radio or at X-ray or GeV energies. These have been dubbed the ``dark accelerators''. 

Being located in the galactic plane, these sources are however generally along the line of sight of molecular clouds. Some, like the HEGRA source in Cygnus (TeV J2032+4131; see Mart\'i et al. 2007; Butt et al. 2008a, b), may be associated with massive star-forming regions, yet are not extended or associated with known active objects like pulsars, supernovae or even compact clusters of massive stars with strong stellar winds like Wes 2 (Aharonian et al. 2007). The spectral information does not help, since the \gr~spectrum of the Cygnus source is a power-law (Albert et al. 2008), as may be the case for IC443 and W28. A background AGN is a possibility, as has been invoked for that source (Butt et al. 2008b). Others, like HESS J1841-055, are extended but seem on the contrary to have too many active objects in their vicinity (several pulsars and SNRs, one high-mass X-ray binary, etc.), although of course a combined action of several local ``accelerators'' cannot be excluded (Aharonian et al. 2008a).

One should however keep in mind that the ``hunt for supernovae'', after all one of the most obvious particle ``accelerators'', is far from over. An ongoing VLA search for non-thermal radio sources in the galactic plane has revealed many new candidate SNRs (e.g., Tian et al. 2008). Other kinds of SNR may not exhibit the usual signposts of their existence: for instance, a neutron star has recently been found near the center of the Carina nebula (Hamaguchi et al. 2009), but so far no SNR has been found to be associated with it. Another example is the famous Eagle nebula M16, in which the possible existence of an SNR has been inferred from unusual properties of its dust emission (Flagey et al. 2009). Note however that, in these two examples, no TeV emission has been detected, in spite of the fact that these regions are typically associated with a large amount of molecular mass ($\sim 10^5 - 10^6 M_\odot$). This may be a new problem, namely the {\it absence} of \gr~emission, hence of significant LCR (hadrons or leptons) acceleration, in these regions.


\section{Concluding remarks}

After {\sl COS-B}, {\sl Compton GRO/EGRET}, and now awaiting relevant results from {\sl Fermi/LAT}, ground-based \u{C}erenkov telescopes have established that massive star-forming regions hosting known SNRs (presumably from SNII, core-collapse supernovae), i.e., SNOBs, are a class of GeV and/or TeV \gr~sources. Thus, local cosmic-ray acceleration, via the so-called ``diffusive shock acceleration'' model, is expected to take place. The presence of associated molecular clouds as targets is the only way to reveal the interaction of the high-energy ($>$ GeV-TeV) component of these cosmic rays via subsequent \pz~decay. However, whereas the presence of extended TeV \gr~emission spatially correlated with CO molecular clouds supports such a mechanism, there is no spectral proof yet that this is the case. 

We have proposed a new tool to investigate this interaction, by looking for enhanced ionization of the TeV-bright molecular clouds by the low-energy ($\sim 10-100$ MeV) component of the same cosmic rays. The results of the relevant observations in the millimeter range are pending. If successful, novel constraints on the propagation of locally accelerated cosmic rays in the vicinity of SNR shocks will be obtained.




\end{document}